\documentclass[12pt]{article}
\usepackage{epsfig}

\newcommand{\be}{\begin{equation}}
\newcommand{\ee}{\end{equation}}
\newcommand{\bea}{\begin{eqnarray}}
\newcommand{\eea}{\end{eqnarray}}

\begin{document}
\begin{titlepage}
%\begin{flushright}
%{\tt    \\preprint n.
%    \\
%    gr-qc/0108050}
 %\end{flushright}
\bigskip%\vskip3mm
\begin{center}

{\bf{\LARGE Quantum gravitational corrections to black hole geometries}}

\bigskip
\bigskip
\bigskip
M.\ Cavagli\`a$\ ^{a,b}$ \footnote{\sc cavaglia@mitlns.mit.edu},
A.\ Fabbri$\ ^c$ \footnote{\sc fabbria@bo.infn.it}
\end{center}
\bigskip
\footnotesize
\noindent
a) Center for Theoretical Physics, Massachusetts Institute of Technology,
77 Massachusetts Avenue, Cambridge MA 02139-4307, USA
\newline
b) INFN, Sede di Presidenza, Roma, Italy
\newline
c) Dipartimento di Fisica dell'Universit\`a di Bologna and
INFN sezione di Bologna, Via Irnerio 46, 40126 Bologna, Italy
\bigskip%\vskip2mm
\bigskip%\vskip2mm
\begin{center}
{\bf Abstract}
\end{center}
We calculate perturbative quantum gravity corrections to eternal
two-dimensional black holes. We estimate the leading corrections to the
$AdS_2$ black hole entropy and determine the quantum modification of
$N$-dimensional Schwarzschild spacetime.
\bigskip
% PACS:
% Keywords:
\end{titlepage}
\newpage
%
%%%%%%%%%%%%%%%%%%%%% INTRODUZIONE%%%%%%%%%%%%%%%%%%%%%%%%%%%%%%%%%%%%%%%%%%%%%%%%%
\section{Introduction}
In recent years, investigation of two-dimensional black holes has earned a
great deal of attention because of its relevance to the study of classical and
quantum properties of (higher-dimensional) black holes, branes, string theory,
and gravitational collapse. (For recent reviews, see, e.g.,
\cite{Strobl:1999wv} and \cite{Nojiri:2001ja}.)

Black holes in two dimensions are usually described by dilaton gravity
theories. In absence of matter fields\footnote{Here, {\it matter field}
means any field but the gravitational field.} two-dimensional dilaton
gravity describes eternal black holes. Matterless dilaton gravity enjoys a
number of interesting properties \cite{Cavaglia:1999xj}: a) The field
equations are completely integrable; b) All solutions are static, i.e.,
depend on a single coordinate (generalized Birkhoff Theorem); c) A local
gauge-invariant integral of motion $M$ exists; d) The moduli space of the
theory is one-dimensional and is described by the modular parameter $M$.
Properties a)-d) imply that matterless two-dimensional dilaton gravity is
a topological theory with no propagating degrees of freedom. Owing to the
Birkhoff theorem the physics on the gauge shell is completely determined
by degrees of freedom on the spacetime boundary. The bulk field modes
describe pure gauge degrees of freedom: All physical information is
contained in the boundary, where the conserved charge is defined. If
matter fields are present properties a)-d) do not generally hold.

It is worth noticing that a nontrivial relation between local pure gauge
degrees of freedom on the bulk and topological degrees of freedom on the
boundary has been suggested for topological theories
\cite{Carlip:1995gy,Strominger:1998eq}. In particular, Carlip has found an
explicit realization of this relation for three-dimensio\-nal gravity
\cite{Carlip:1995gy}. The idea is that the gauge invariance of
three-dimensional gravity is broken by the presence of the boundary. As a
consequence, pure gauge degrees of freedom may become dynamical, i.e.\ {\em
off-shell}, and induce physical degrees of freedom on the boundary. A similar
result may be valid for two-dimensional matterless dilaton gravity.

At the quantum level, two main different lines of research have been pursued in
the literature. In the {\em semiclassical approach} matter fields are quantized
and gravity is classical. The source term in the Einstein equations is the
vacuum expectation value of the stress energy tensor of matter fields evaluated
on the classical geometry. Perturbative quantum corrections to the classical
metric are obtained by solving the equations at the lowest order in $\hbar$.
(See for instance \cite{BirrellDavies,Frolov:1996hd}.) Following Hawking's
discovery that black holes emit thermal radiation \cite{Hawking:1975sw}, a
large number of semiclassical toy models have been used to describe the
evolution of evaporating black holes. (See, e.g.,
\cite{Harvey:1992xk}-\cite{Thorlacius:1995ip}.) In the {\em quantum gravity
formalism} the gravitational sector of the theory is quantized. This approach
has been carried on by a number of authors
\cite{Hirshfeld:2000xm}-\cite{Cavaglia:2000zu}, with seemingly contradictory
results. On one hand, Poisson sigma models \cite{Hirshfeld:2000xm} and
first-order path integral quantization \cite{Kummer:1997hy} of dilaton gravity
in absence of matter fields seem to indicate that quantum corrections to the
effective action {\em on-shell} are zero at all orders of perturbation theory.
On the other hand, nonlinear sigma model methods
\cite{Kazakov:1994ha,Cavaglia:2000zu}, quantization of gravity as an effective
field theory \cite{Donoghue:1994dn,Donoghue:1994eb}, and Matrix theory 
\cite{Becker:1999uz} lead to nonzero quantum deformations of the classical
metric. Although a comparison between the different approaches is difficult for
technical reasons, the origin of the disagreement is most likely due to the
different treatment of boundary terms: If pure gauge degrees of freedom become
dynamical, as suggested in \cite{Carlip:1995gy,Strominger:1998eq}, the quantum
corrections to the effective action are generally nonzero already at one-loop
\cite{Kantowski:1992dv}-\cite{Elizalde:1994qq}. (See also
\cite{Nojiri:2001ja}.)

In this paper we follow a third approach to the quantization of two-dimensional
black holes. Since our aim is to compute pure quantum gravity corrections to
the classical geometry of (eternal) black holes, i.e.\ to the mass $M$, we
quantize the moduli space of the theory. The moduli space approximation is a
standard procedure that can be used when there is a continuous family of static
solutions. (See, e.g., \cite{Khoury:2001wf} and references therein.) The action
for the moduli space is obtained by substituting the static solutions in the
original action with the moduli represented as spacetime fields. The moduli
action for two-dimensional black holes coincides with the nonlinear sigma model
formulation of two-dimensional dilaton gravity \cite{Cavaglia:1999xj}. Thus the
quantization of two-dimensional black holes geometries is formally equivalent
to the quantization of the string in curved spacetime \cite{GSW}. By applying
the perturbative quantization algorithm described in Ref.\
\cite{Cavaglia:2000zu} we evaluate pure quantum gravity corrections to the
geometry of eternal dilatonic black holes.  Quantum corrections to the
classical solutions are obtained by a perturbative expansion on powers of the
curvature. Quantum gravitational effects vanish in the limit of large ADM
mass and in the asymptotic region far away from the black hole horizon where
the black hole behaves classically. Quantum effects become instead significant
at finite distances from the black hole horizon. This leads to a modification
of the physical quantities which are associated to the classical geometry.

The structure of the paper is as follows. In the next section we briefly review
the classical theory and find the moduli action for two-dimensional eternal 
black holes The latter is quantized in Section 3. As two illustrative examples
we estimate the leading quantum gravity corrections to the Bekenstein-Hawking
entropy of the $AdS_2$ black hole and to $N$-dimensional spherically symmetric
gravity. In Section 4 we state our conlusions.
\section{Classical setting and moduli space}
In the Schwarzschild gauge two-dimensional eternal black holes are
described by the metric
\be\label{sol}
ds^2_{(2)}=-\left[N(x)-M\right]dt^2
+ \left[N(x)-M\right]^{-1}dx^2\,,
\ee
where $M$ is a constant parameter (modulus) and $N(x)$ is a function of the
spacetlike coordinate $x$. The simplest theory that admits Eq.\ (\ref{sol}) as
general solution is
\be
\label{act}
S_G=\int d^2x \sqrt{-g}\left[ \phi R + V(\phi)\right]\,.
\ee
$R$ is the two-dimensional Ricci scalar constructed from the metric
$g_{\mu\nu}(x)$ and $\phi$ is a scalar field which is usually called
``dilaton'' in the literature. In light cone conformal coordinates the
general solution of Eq.\ (\ref{act}) is \cite{Filippov:1996ye,Cavaglia:1999xj}
\be\label{sol-conf}
ds^2_{(2)}\equiv g_{\mu\nu}\,dx^\mu dx^\nu=4\left[N(\phi)-M\right]dudv\,,
\ee
where $M$ is a constant (modulus) and $\phi$ is a function of $x\equiv
u+v$ defined by the differential equation
\be\label{defphi}
{dx\over d\phi}={1\over N(\phi)-M}\,,\qquad
N(\phi)=\int^{\phi} d\phi ' V(\phi ')=\phi^b\,.
\ee
Equation (\ref{sol-conf}) coincides, in the Schwarzschild gauge, with Eq.\
(\ref{sol}). In the following we will restrict attention to the power-law
dilaton potential
\be\label{pot}
V(\phi)=b \phi^{b-1}\,,
\ee
where $b$ is a positive real number. The constant parameter $M$ is proportional
to the ADM mass. The geometry is singular at $\phi=0$ and asymptotically flat
as $\phi\to\infty$. The black hole event horizon is located at
$\phi=\phi_h=M^{1/b}$.

The action (\ref{act}) can also be used to describe the two-dimensional
effective theory of $N$-dimensional spherically symmetric gravity. Setting
\be\label{schwarz}
ds^2_{(N)}=\phi^{-b}g_{\mu\nu}dx^\mu dx^\nu+
\phi^{2(1-b)}d\Omega^2_{N-2}\,.
\ee
the general solution in the Schwarzschild gauge reads \cite{Cavaglia:2000zu}
\be
\label{nsch} ds^2_{(N)}=-(1-J/r^{N-3})dt^2
+\frac{dr^2}{(1-J/r^{N-3})}+r^2d\Omega^2_{N-3}\,,
\ee
where the parameter $J$ is related to the ADM mass by
\be
\label{admm}
J=\frac{16\pi l_{pl}^{N-2}}{(N-2)V_{N-2}}M_{ADM}\,,
\ee
and $V_{N-2}$ is the volume of the $(N-2)$-dimensional unit sphere. ($l_{pl}$
is the $N$-dimensional Planck length.)

In Ref.\ \cite{Cavaglia:1999xj} it is shown that the theory (\ref{act}) can be
formulated as a nonlinear sigma model. We introduce the new field
\be\label{mass}
M(\phi,g_{\mu\nu})=N(\phi)-\partial_\mu\phi\partial^\mu\phi\,.
\ee
$M(\phi,g_{\mu\nu})$ is invariant under gauge transformations and is locally
conserved on-shell, where it coincides with the parameter $M$ in Eq.\
(\ref{sol}). Neglecting inessential surface terms Eq.\ (\ref{act}) can be cast
in the form
\be\label{sigma}
S_{\sigma}={1\over 2}\int d^2x\,G_{ij}(\xi)\partial_\mu\xi^i\partial^\mu\xi^j\,,
\ee
where $\xi^i=(M,\phi)$ and the metric of the target space is
\be\label{target}
G_{ij}={1\over
N(\xi^1)-\xi^0}\Omega_{ij}\,,\qquad\Omega_{ij}=\left(\matrix{0&1\cr
1&0}\right)\,.
\ee
$G_{ij}$ is singular at the black hole horizon and asymptotically flat for
$\phi\to\infty$. Equation (\ref{sigma}) coincides with the moduli action of the theory.
The latter is defined as [remember Eq.\ (\ref{sol-conf})]
\be
\label{act-moduli}
S_{moduli}\equiv S_{G}[g_{\mu\nu}=\rho(x)\eta_{\mu\nu}]\,,
\ee
where $\rho(x)=N[\phi(x)]-M(x)$. Using Eq.\ (\ref{mass}) in Eq.\
(\ref{act-moduli}) we find $S_{moduli}=S_{\sigma}$. The moduli action
(\ref{sigma}) is our starting point in the calculation of quantum corrections
to black hole geometries.
\section{Quantization}
The moduli action can be quantized perturbatively by expanding the metric of
the target space in Riemann normal coordinates $\zeta$
\cite{Alvarez-Gaume:1981hn}\footnote{For the derivation and a detailed
discussion of Riemann normal coordinates see e.g.\ Ref.\ \cite{Petrov}.}
\be\label{Riemann}
G_{ij}(\xi)=G_{ij}[\xi(0)]-{1\over 3}R_{ikjl}[\xi(0)]\zeta^k
\zeta^l-{1\over 3!}R_{ikjl;m}[\xi(0)]\zeta^k \zeta^l \zeta^m+\dots\,,
\ee
where $R_{ikjl}$ is the Riemann tensor. The Riemann coordinates are (up to
second order in the Riemann expansion)
\be\label{Riemann-coord}
\xi^i=\xi^i(0)+\zeta^i-{1\over
2}\Gamma^{i}{}_{jk}[\xi(0)]\zeta^j\zeta^k-{1\over
{3!}}\Gamma^i_{jkl}[\xi(0)]\zeta^j\zeta^k\zeta^l + \dots\,,
\ee
where
\be
\Gamma^i_{jkl}\equiv {1\over 3}P\left( \Gamma^i_{jk,l}-2\Gamma^i_{\alpha
j} \Gamma^{\alpha}_{kl}\right) \> ,
\ee
and $P$ indicates the sum of terms obtained by permuting the subscripts
cyclically. The corrections to the fields are defined as
\be\label{delta-fields}
\delta\xi^i\equiv\zeta^i-{1\over
2}\Gamma^{i}{}_{jk}[\xi(0)]\zeta^i\zeta^k -{1\over {3!}}
\Gamma^i_{jkl}[\xi(0)]\zeta^j\zeta^k\zeta^l \dots
\ee
The perturbative action follows by inserting the Riemann expansion
(\ref{Riemann}) in Eq.\ (\ref{sigma}):
\be\label{act-pert}
S_{\sigma,pert}={1\over 2}\int d^2x\,\left[\partial_\mu
\zeta^i\partial^\mu\zeta^j\Omega_{ij}+g(\zeta)\epsilon_{ij}\epsilon_{kl}\partial_\mu
\zeta^i\partial^\mu \zeta^k \zeta^j \zeta^l\right]\,,
\ee
where $\epsilon_{ij}$ is the two-dimensional Levi-Civita tensor density and
$g(\zeta)$ is the ($\zeta$-dependent) coupling constant
\be
g(\zeta)=\sum_{m,n=0} g_{mn}(\phi_0,M_0)
(\zeta^1)^m(\zeta^0)^n\,.
\ee
Due to the curvature singularity of the target metric (\ref{target}), all
terms in the coupling constant expansion $g(\zeta)$ diverge on the black hole
horizon; the latter corresponding to the strong coupled regime of the theory.
Therefore the black hole horizon cannot be described by the perturbative
expansion (\ref{act-pert}).

In Eq.\ (\ref{act-pert}) we have rescaled the fields $\zeta^i$ by a factor
\be\label{rescaling}
\zeta\to\zeta'=\zeta[N(\phi_0)-M_0]^{-1/2}\,,
\ee
and have dropped the primes for convenience. At the leading order in the
perturbative region ($\phi_0\to\infty$) the coupling constants $g_{mn}$ are
\be\label{gmn}
g_{mn}\approx\phi_0^{-bn/2+m(b/2-1)-1}\,.
\ee
Equation (\ref{act-pert}) describes an interacting string which propagates in a
curved two-dimensional target space. The first-order coupling constant,
$g_{00}\approx\phi_0^{-1}$, can be interpreted as the (coordinate dependent)
coupling constant of the theory (\ref{act}). Thus the perturbative expansion
(\ref{act-pert}) is both a weak-coupling expansion in terms of the
coordinate-dependent gravitational coupling of the model and an expansion near
the boundary at $\phi_0\to\infty$.

Using Eq.\ (\ref{target}), Eq.\ (\ref{delta-fields}) and Eq.\ (\ref{rescaling})
the asymptotic corrections to the physical fields can be expressed as functions
of $\phi_0$ and $\zeta^i$. For instance, at the third order in the Riemann
expansion we have
\bea\label{delta-M}
\delta M\equiv\delta\xi^0\approx\phi_0^{b/2}\zeta^0
&+&A(\zeta^0)^2+
B\phi_0^{-b/2}(\zeta^0)^3+
C\phi_0^{b/2-1}(\zeta^0)^2\zeta^1+\nonumber\\
&+&D\phi_0^{-b}(\zeta^0)^4+
E\phi_0^{-1}(\zeta^0)^3\zeta^1+
F\phi_0^{b-2}(\zeta^0)^2(\zeta^1)^2+\nonumber\\
&+&\dots
\eea
and
\bea\label{delta-phi}
\delta\phi\equiv\delta\xi^1\approx\phi_0^{b/2}\zeta^1
&+&A'\phi_0^{b-1}(\zeta^1)^2+
B'\phi_0^{3b/2-2}(\zeta^1)^3+
C'\phi_0^{b/2-1}(\zeta^1)^2\zeta^0+\nonumber\\
&+&D'\phi_0^{2b-3}(\zeta^1)^4+
E'\phi_0^{b-2}(\zeta^1)^3\zeta^0+
F'\phi_0^{-1}(\zeta^1)^2(\zeta^0)^2+\nonumber\\
&+&\dots\,,
\eea
for $M$ and $\phi$, respectively.

>From Eq.\ (\ref{act-pert}) it is straightforward to read the Feynman rules of
the theory. The free propagator is antidiagonal in the fields
\be\label{free}
\langle\zeta^i(x_1)\zeta^j(x_2)\rangle=-\,\Omega^{ij}\,{1\over 4\pi}\,
\ln{\left[(x_1-x_2)^2\right]}\,.
\ee
Graphically:
\be
\epsfig{file=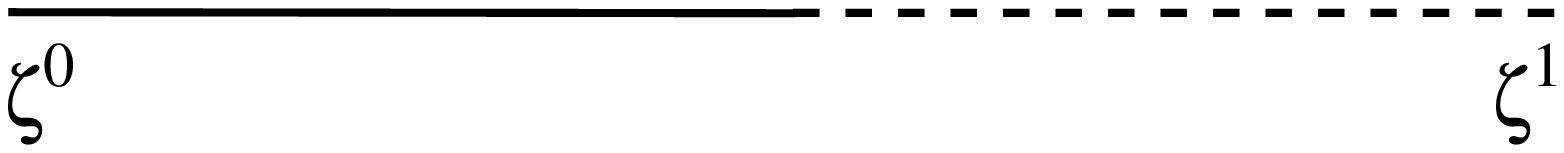, width=2truein}
\ee
For each interaction term in Eq.\ (\ref{act-pert}) we have a vertex
\be\label{vertex}
\epsfig{file=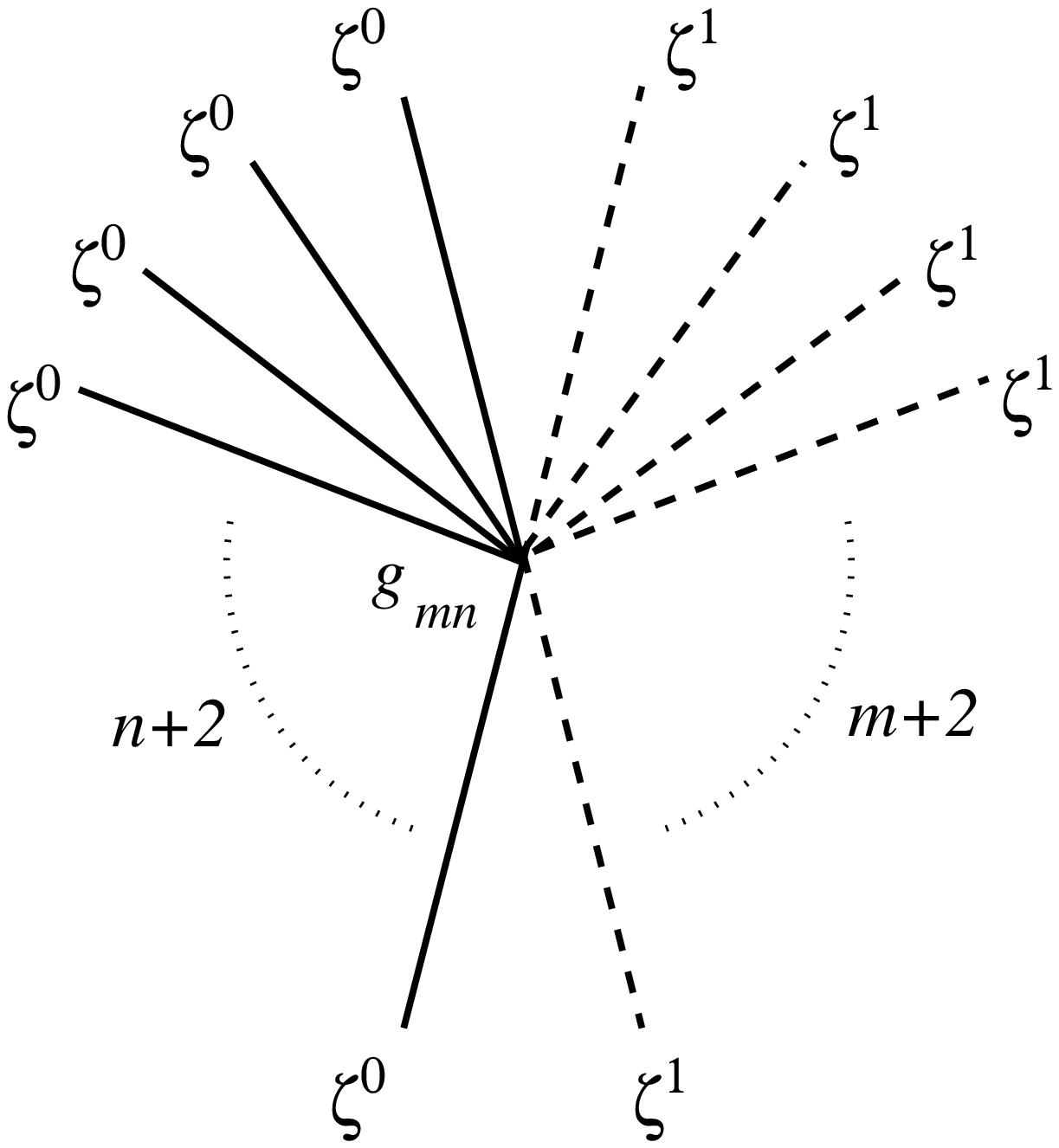, height=2truein}
\ee
which is generated by the interaction term of order $k=n+m+1$ in the Riemann
expansion.

Classical conformal invariance of the moduli action is broken at quantum level.
In the conformal gauge in $d=2-\epsilon$ dimensions the (renormalized) moduli
action is
\be\label{action-ren}
S=\int d^dx\, e^{-\epsilon\rho/2}\,{\cal L}_{ren}\,+S_{\rho}+S_{ghost}\,,
\ee
where $S_{\rho}$ is the effective action due to the Weyl anomaly and ${\cal
L}_{ren}$ is renormalized Lagrangian. At one-loop the breaking of conformal
invariance leads to the following anomalous vertices
\be\label{vertex-anom}
\epsfig{file=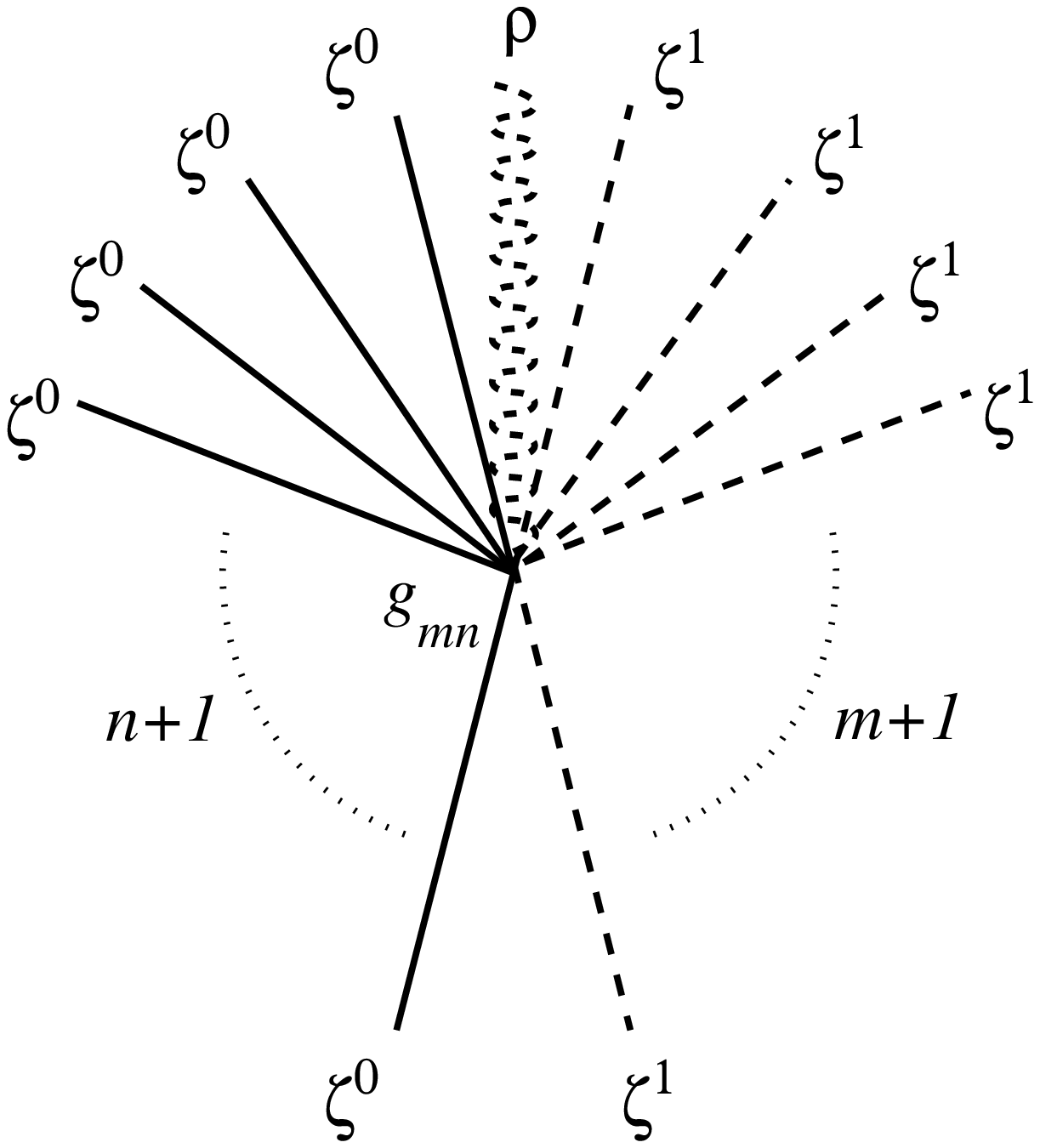, width=2truein}
\ee
We can calculate the corrections to the free propagator due to the interaction
vertices of the string (\ref{vertex}) and (\ref{vertex-anom}). At the first
order in the Riemann curvature expansion the topology of the interaction
implies that the propagator is antidiagonal at any loop. The one-loop
correction to the two-point Green function is given by the Feynman graph
\be\label{two-point}
\epsfig{file=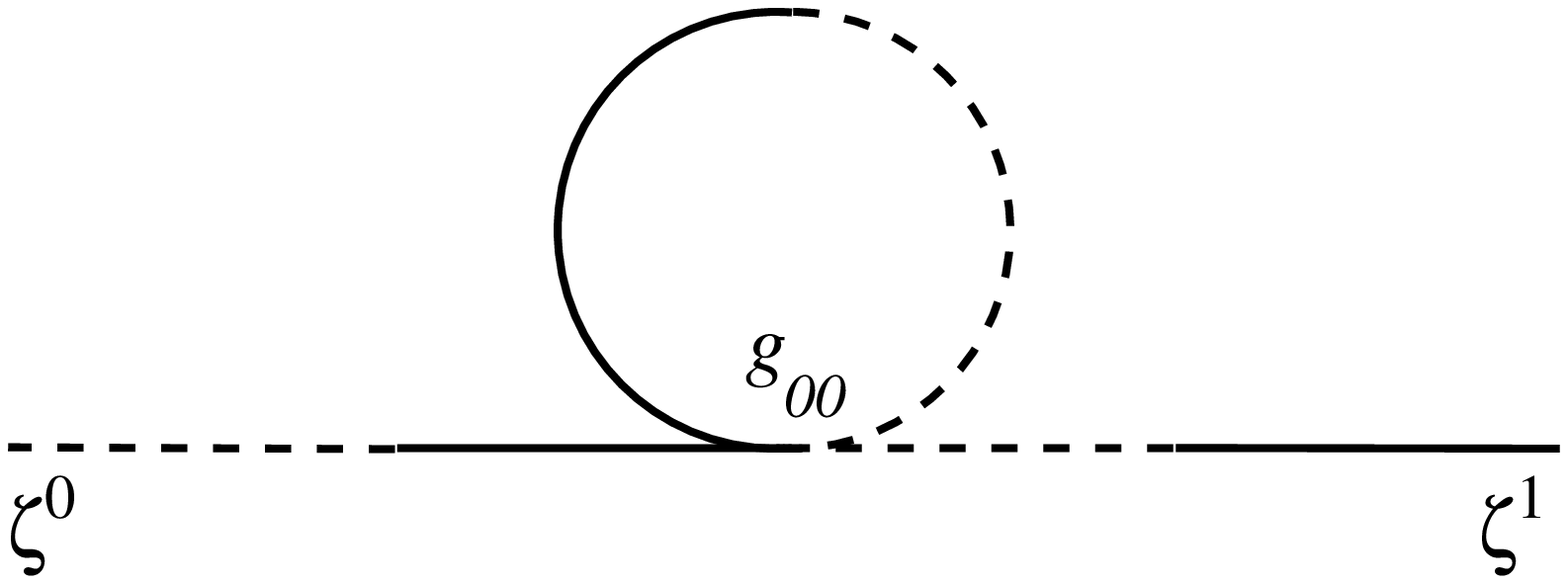, width=2truein}
\ee
and by the two anomalous graphs
\be\label{two-point-anom-fo}
\epsfig{file=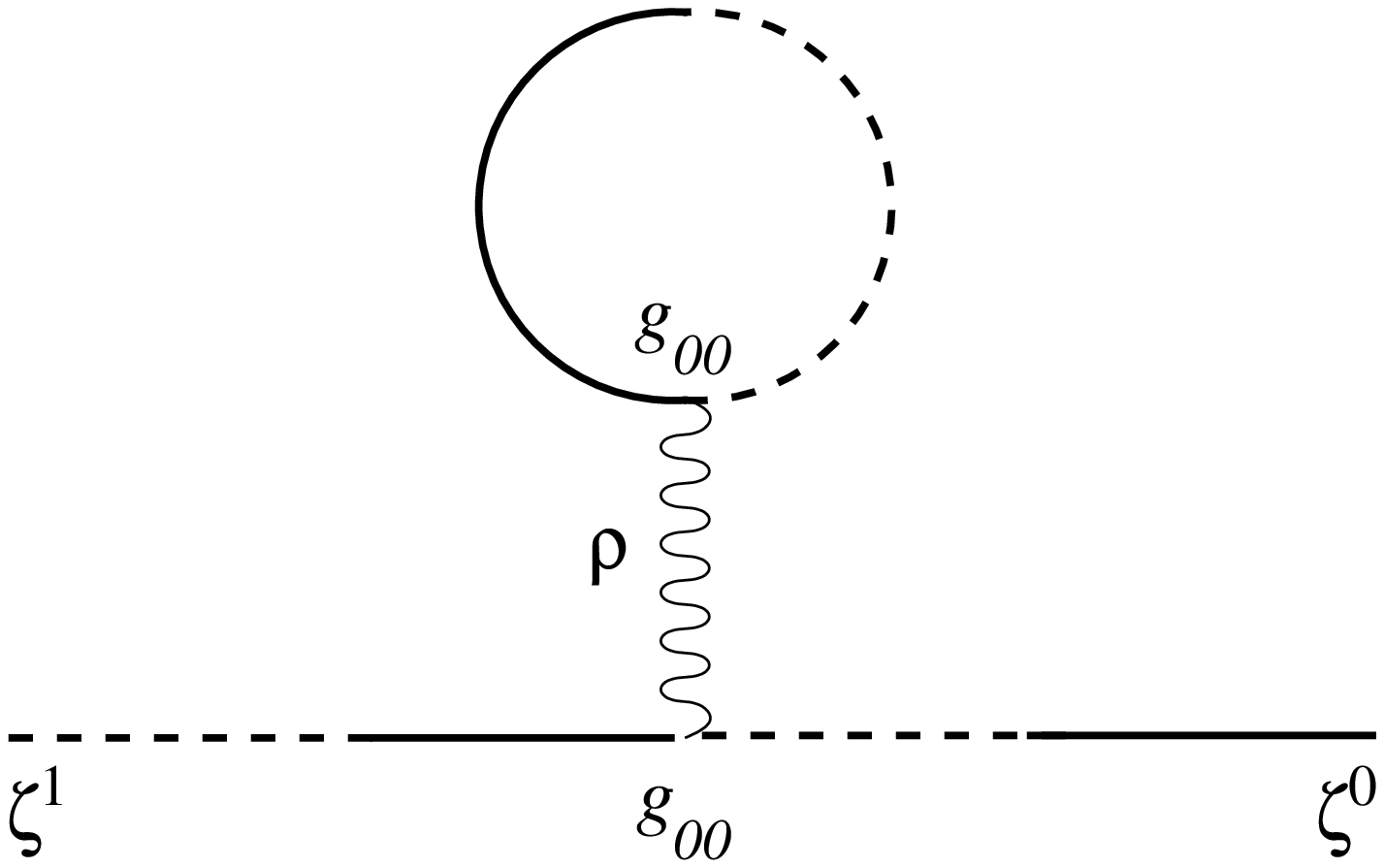, width=2truein}\qquad
\epsfig{file=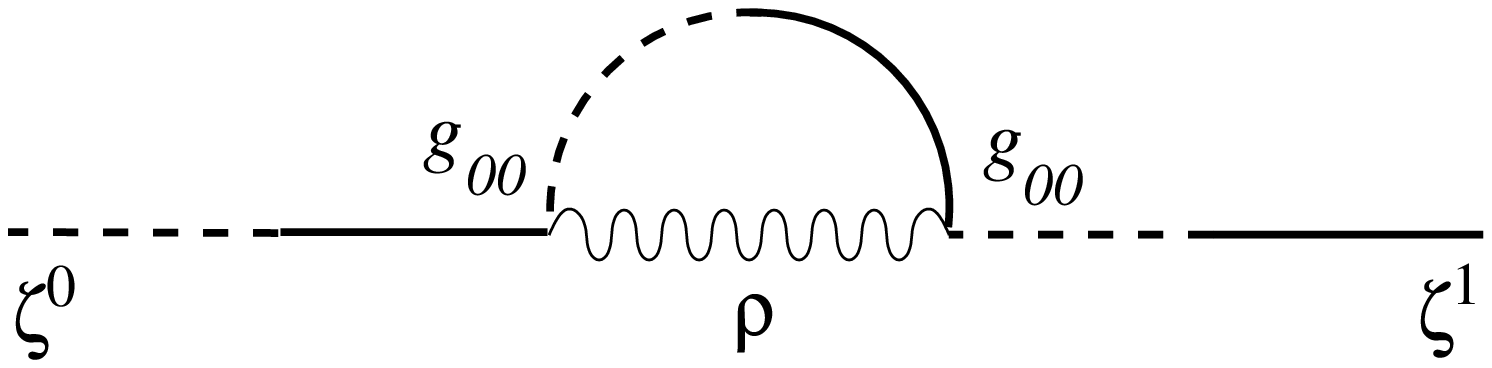, width=2truein}
\ee
The two anomalous graphs are subdominant with respect to the contribution in
Eq.\ (\ref{two-point}). Following Ref.\ \cite{Cavaglia:2000zu} the one-loop
renormalized two-point function is
\be\label{two-point-ren}
\langle\zeta^i(x_1)\zeta^j(x_2)\rangle=-\Omega^{ij}{1\over
4\pi}\left[1-{g_{00}\over 4\pi}\ln\left({\mu\over
m}\right)^2+O(g_{00}^2)\right] \ln\left[(x_1-x_2)^2\right]\,.
\ee
(The leading nonzero diagonal contribution to the propagator is obtained at two
loops and at the third order in the Riemann curvature expansion.) It is
straightforward to evaluate the contributions to higher order Green functions.
Here we will give the result for three and four-point functions at one-loop
and up to the third order in the Riemann expansion. The leading contribution to
the three-point Green functions is given at one-loop by Feynman graphs with a
single five-point vertex:
\be\label{three-point}
\langle\zeta^0(x_1)\zeta^0(x_2)\zeta^1(x_3)\rangle\approx
O(\phi_0^{b/2-2})\,, \quad
\langle\zeta^0(x_1)\zeta^1(x_2)\zeta^1(x_3)\rangle\approx
O(\phi_0^{-b/2-1})\,.
\ee
(The leading non-vanishing contribution to the three-point Green functions with
identical external legs is obtained at two-loops and at the fourth order in the
Riemann expansion.) At one-loop the nonvanishing four-point Green functions are
\bea\label{four-point}
&&\langle\zeta^0(x_1)\zeta^0(x_2)\zeta^0(x_3)\zeta^1(x_4)\rangle\approx
O(\phi_0^{b-3})\,,\nonumber\\
&&\langle\zeta^0(x_1)\zeta^0(x_2)\zeta^1(x_3)\zeta^1(x_4)\rangle\approx
O(\phi_0^{-1})\,,\\
&&\langle\zeta^0(x_1)\zeta^1(x_2)\zeta^1(x_3)\zeta^1(x_4)\rangle\approx
O(\phi_0^{-b-1})\,.\nonumber
\eea
The (one-loop) perturbative quantum corrections to the classical geometry
follow from Eqs.\ (\ref{delta-M})-(\ref{delta-phi}) and Eqs.\
(\ref{two-point-ren})-(\ref{four-point}). At the third order in the Riemann
expansion we find
\bea\label{delta-green}
\langle\delta M\rangle_{1-loop}&\approx&
C\phi_0^{b/2-1}\langle(\zeta^0)^2\zeta^1\rangle+
E\phi_0^{-1}\langle(\zeta^0)^3\zeta^1\rangle+
F\phi_0^{b-2}\langle(\zeta^0)^2(\zeta^1)^2\rangle\nonumber\\
&\approx& O(\phi_0^{b-3})\,,\\\nonumber\\
\langle\delta\phi\rangle_{1-loop}&\approx&
C'\phi_0^{b/2-1}\langle(\zeta^1)^2\zeta^0\rangle+
E'\phi_0^{b-2}\langle(\zeta^1)^3\zeta^0\rangle+
F'\phi_0^{-1}\langle(\zeta^1)^2(\zeta^0)^2\rangle\nonumber\\
&\approx& O(\phi_0^{-2})\,.
\eea
Finally, from the free perturbative action we have
\be\label{delta-green-two}
\langle\delta M\delta\phi\rangle_{1-loop}\approx O(\phi_0^{b})\,.
\ee
The leading contribution to $\langle\delta M\delta\phi\rangle_{1-loop}$ is
given at the zeroth order in the Riemann expansion by the term
$\phi_0^{b}\zeta^0\zeta^1$. Nonzero values of $\langle\delta
M\rangle_{1-loop}$ and $\langle\delta\phi\rangle_{1-loop}$ are first generated
at the second order in the Riemann expansion when $\zeta^3$ terms appear in the
expansions (\ref{delta-M}) and (\ref{delta-phi}) and three-point Green
functions are non-vanishing. The corrections to $\langle\delta
M\rangle_{1-loop}$ and $\langle\delta\phi\rangle_{1-loop}$ that are generated
by the third order term in the Riemann expansion are of the same order of the
corrections generated by the second order term. This is due to the presence of
the $(\zeta^0)^2(\zeta^1)^2$ term in the expansions (\ref{delta-M}) and
(\ref{delta-phi}) which is nonzero at tree level. Higher order corrections that
follow from higher order terms in the Riemann expansion are subleading.

Using Eqs.\ (\ref{delta-green})-(\ref{delta-green-two}) we can compute the
corrections to the classical metric (\ref{sol}). In the $(t,\phi)$ gauge we have
\be\label{delta-g} \delta g_{\mu\nu}\approx
\left(\matrix{\phi_0^b\left[{b\delta\phi\over\phi_0}-{\delta
M\over\phi_0^b}\right]&0\cr 0&{1\over
\phi_0^b}\left[{b\delta\phi\over\phi_0}-{\delta
M\over\phi_0^b}\right]}\right)\,. \ee
Since no cross term $\delta M\delta\phi$ appears in Eq.\ (\ref{delta-g}) the
leading corrections to the classical geometry are generated at the second order
in the Riemann expansion. We shall see below that the leading quantum
corrections to the $N$-dimensional Schwarzschild black hole are given instead
by Eq.\ (\ref{delta-green-two}).

At one-loop the leading quantum corrections to $g_{\mu\nu}$ are
\be\label{delta-g-green}
\langle\delta g_{\mu\nu}\rangle_{1-loop}=\left(\matrix{O(\phi_0^{b-3})&0\cr
0&O(\phi_0^{-b-3})}\right)\,.
\ee
These corrections modify the asymptotic structure of the spacetime and have
important consequences for physical quantities which are defined on the
boundary. The quantum-corrected line element at one-loop is
\be\label{sol-one-loop}
ds^2_{1-loop}=\left[\phi^b-M+O(\phi^{b-3})\right]dt^2-\left[\phi^b-M+
O(\phi^{b-3})\right]^{-1}d\phi^2\,. \ee
This metric is a solution of the effective action
\be
\label{act-eff}
S_{eff}=\int d^2x\sqrt{-g}\left[\phi R+V_{eff}(\phi)\right]\,,
\ee
where
\be\label{pot-eff}
V_{eff}(\phi)=b\phi^{b-1}+O(\phi^{b-4})\,.
\ee
Let us now discuss a couple of specific models.
\subsection{$b=2$: AdS$_2$}
The case $b=2$ is rather interesting. If $b=2$ Eq.\ (\ref{act}) describes the
Jackiw-Teitelboim model \cite{JT} and the classical metric is anti-de Sitter
(AdS). The latter describes the near-horizon limit of four-dimensional
Reissner-Nordstr\"om black holes close to extremality
\cite{Spradlin:1999bn}-\cite{Fabbri:2001xh}. Thanks to the AdS/CFT
correspondence, dilaton gravity on AdS$_2$ is dual to a conformal field theory
defined on the AdS$_2$ boundary: The gravitational dynamics can be described by
the microscopic degrees of freedom of the CFT
\cite{Cadoni:2000kr,Cadoni:2001fq}. In particular, the statistical entropy of
two-di\-men\-sio\-nal AdS black holes can be derived by counting the
microstates on the one-dimensional (timelike) boundary of AdS$_2$
\cite{Cadoni:1999sg,Catelani:2000gn,Navarro-Salas:2000up}. The result is
\be\label{th-entropy-ads}
S_{AdS}=2\pi\sqrt{M}\,.
\ee
In Ref.\ \cite{Cadoni:1999ja} Cadoni and Mignemi have generalized the previous
calculation to models with asymptotic dilatonic potential
\be\label{JT-generalized}
V(\phi)=2\phi+O(\phi^{-2})\,.
\ee
They find that the black hole entropy coincides with the AdS entropy at the
leading order. So the statistical entropy of the two-dimensional AdS black hole
($b=2$) is not modified by (perturbative) one-loop quantum gravity effects at
the leading order. If we trust the thermodynamical derivation of the black hole
entropy \cite{Wald:1993nt,Iyer:1994ys,Myers:1994sg} beyond the semiclassical
regime, the one-loop thermodynamical entropy of the AdS black hole is, in the
large mass limit,
\be\label{th-entropy-oneloop}
S_{AdS,1-loop}=2\pi\sqrt{M}+O(M^{-1})\,.
\ee
Therefore, quantum gravity corrections to the thermodynamical entropy are
subdominant to the logarithmic corrections that follow from the Cardy formula
\cite{Kaul:2000kf,Carlip:2000nv}.
\subsection{$b=N-3/(N-2)$: $N$-dimensional Schwarzschild}
Let us now discuss the quantum corrections to the $N$-dimensional
Schwarz\-schild black hole. Because of the presence of the two-dimensional
conformal factor $[\phi(x)]^{-b}$ in Eq.\ (\ref{schwarz}), the expectation
values of $g_{tt}^{(N)}$ and $g_{rr}^{(N)}$ involve the term $\langle \delta M
\delta \phi \rangle$. The latter gives the leading quantum corrections in the
perturbative limit $\phi_0\to\infty$. The quantum-corrected geometry is
\be
\label{qcsg}
ds^2=-\left[1-{J\over r^{N-3}}\left(1+O\left({1\over
r}\right)\right)\right]dt^2
+{dr^2\over \displaystyle\left[1-{J\over r^{N-3}}\left(1+O\left({1\over
r}\right)\right)
\right]}+r^2d\Omega^2_{N-2}\,.
\ee
Higher order corrections in the Riemann expansion are subleading with respect
to the $O(1/r)$ terms in Eq.\ (\ref{qcsg}). We have evaluated one-loop
corrections up to the third order. Second- and third-order quantum effects
contribute to $\langle g_{tt}\rangle$ and $\langle g_{rr}^{-1}\rangle$ with
terms of order $O(r^{-3N+6})$ and $O(r^{-4N+9})$. The spherical part of the
$(N-2)$-dimensional metric receives corrections of the order $O(r^{-3N+8})$ and
$O(r^{-4N+10})$.

Comparing Eq.\ (\ref{qcsg}) to the classical metric (\ref{nsch}) we can define
an effective ADM mass:
\be \label{qumz}
M_{ADM,eff}=M_{ADM,cl}+O\left(\frac{1}{r}\right)\,.
\ee
Thus the classical ADM mass receives quantum corrections of order $O(1/r)$, in
agreement with the four-dimensional analysis of \cite{Kazakov:1994ha} and
\cite{Donoghue:1994dn}-\cite{Becker:1999uz}. In four dimensions quantum
gravitational effects are formally equivalent to an electric charge:
Asymptotically, Eq.\ (\ref{qcsg}) describes a Reissner-Nordstr\"om black hole.
For $N>4$ the $O(r^{-N+2})$ term in Eq.\ (\ref{qcsg}) dominates over the
Reissner-Nordstr\"om charge term of order $O(r^{-2(N-3)})$ \cite{Myers:1986un}.
\section{Conclusions}
In this paper we have evaluated the leading quantum gravitational corrections
to eternal two-dimensional black hole spacetimes. We have found that quantum
gravitational corrections to the classical geometry are finite and nonzero. Let
us stress that the existence of nonzero quantum corrections to classical
geometries may be crucial to understand unsolved issues in quantum gravity such
as black hole evaporation, loss of coherence and/or gravitational collapse.

Our results have been obtained by quantizing the moduli action of
two-dimensional dilaton gravity. The action for the moduli fields of eternal
black holes is described by a conformal nonlinear sigma model with a fixed
target metric. The fields are the dilaton and the spacetime dependent modulus
$M$. The sigma model describes a two-dimensional string propagating in a
two-dimensional curved spacetime. Hence, the theory can be quantized
perturbatively by expanding the metric of the target space in normal Riemann
coordinates. Since the expansion parameter is proportional to the curvature of
the manifold, the theory becomes asymptotically free at large distances from
the black hole horizon(s), where the perturbative regime is valid (weak-coupled
region). Finally, we presented two specific models.  We estimated the leading
corrections to the Bekenstein-Hawking entropy of the $AdS_2$ black hole and the
quantum gravitational modification of $N$-dimensional Schwarzschild black hole.
For the Schwarzschild black hole our results confirm the previous
four-dimensional analysis of \cite{Kazakov:1994ha} and
\cite{Donoghue:1994dn}-\cite{Becker:1999uz}.

Let us conclude this paper by mentioning two possible extensions of our
results. The perturbative approach presented above breaks down on the
black hole horizon where the gravitational theory becomes strong-coupled.
Therefore it can not be used to discuss quantum corrections to the
geometry in the near-horizon region. Possibly, this can be accomplished
either by using duality symmetries of the dilaton gravity theory (see
e.g.\ \cite{Cadoni:2001ew}) or by formulating an alternative sigma model
description with nonsingular target metric. Finally, it would be of
primary importance to explore the effects of the inclusion of matter
fields in the action (\ref{act}), and investigate black hole evaporation
in a full quantum gravitational context.
\section*{Acknowledgements}
We are very grateful to R.\ Balbinot, M.\ Cadoni, J.\ Donoghue, R.\ Jackiw, W.\
Kummer and D.\ Vassilevich for interesting discussions and useful comments.
This work is supported in part by funds provided by the U.S.\ Department of
Energy (D.O.E.) under cooperative research agreement DE-FC02-94ER40818.

\end{document}